\documentclass[%
reprint,
superscriptaddress,
amsmath,amssymb,
prd,
]{revtex4-2}

\usepackage[caption = false]{subfig}

\usepackage{newtxtext,newtxmath}
\usepackage{graphicx}
\usepackage{dcolumn}
\usepackage{bm}
\usepackage{hyperref}

\usepackage{blindtext}

\usepackage{siunitx}

\DeclareSIUnit\Mpc{Mpc}
\DeclareSIUnit\Gpc{Gpc}
\DeclareSIUnit\Gyr{Gyr}

\usepackage{tikz}
\usepackage{pgfplots}
\usepgfplotslibrary{fillbetween}

\definecolor{rosso}{RGB}{220,57,18}
\definecolor{giallo}{RGB}{255,153,0}
\definecolor{blu}{RGB}{102,140,217}
\definecolor{verde}{RGB}{16,150,24}
\definecolor{viola}{RGB}{153,0,153}

\newcommand\aap{A\&A}                
\newcommand\mnras{MNRAS}             
\newcommand\jcap{J.~Cosmology Astropart. Phys.} 
\newcommand\physrep{Phys.~Rep.}      
\newcommand\apjl{ApJ}                
\newcommand\pasj{PASJ}               

\begin{document}

\title{Negative neutrino masses as a mirage of dark energy}
\author{Willem Elbers}\email{willem.h.elbers@durham.ac.uk}
\author{Carlos S. Frenk}
\author{Adrian Jenkins}
\author{Baojiu Li}
\affiliation{Institute for Computational Cosmology, Department of Physics, Durham University, South Road, Durham, DH1 3LE, UK}
\author{Silvia Pascoli}
\affiliation{Dipartimento di Fisica e Astronomia, Universit\`a di Bologna, via Irnerio 46, 40126 Bologna, Italy\\
INFN, Sezione di Bologna, viale Berti Pichat 6/2, 40127 Bologna, Italy}

\date{\today}

\begin{abstract}
The latest cosmological constraints on the sum of the neutrino masses depend on prior physical assumptions about the mass spectrum. To test the accordance of cosmological and laboratory constraints in the absence of such priors, we introduce an effective neutrino mass parameter that extends consistently to negative values. For the $\Lambda$CDM model, we analyze data from \emph{Planck}, ACT, and DESI and find a $2.8-3.3\sigma$ tension with the constraints from oscillation experiments. Motivated by recent hints of evolving dark energy, we analyze the $w_0w_a$ and mirage dark energy models, finding that they favour larger masses consistent with laboratory data, respectively $\sum m_{\nu,\text{eff}} = 0.06_{-0.10}^{+0.15}\,\si{\eV}$ and $\sum m_{\nu,\text{eff}} = 0.04_{-0.11}^{+0.15}\,\si{\eV}$ (both at 68\%).
\end{abstract}

\maketitle

\section{Introduction}

The ability of cosmological surveys to probe the sum of the neutrino masses \cite{lesgourgues06,wong11,abazajian16} provides a unique opportunity to evaluate cosmological models by confronting them with laboratory constraints on this same quantity. The strongest model-independent constraint on the neutrino mass comes from measurements of tritium $\beta$-decay by KATRIN: $m_\beta<\SI{0.45}{\eV}$ (90\% C.L.) \citep{katrin24}. If neutrinos are Majorana particles, searches for neutrinoless double $\beta$-decay are also sensitive to the mass scale, with the strongest limit coming from the KamLAND-Zen experiment \cite{abe24} at $m_{\beta\beta}<0.028-0.122\,\si{\eV}$ (90\%)  \footnote{We clarify that, for the values currently tested by KATRIN, $m_\beta$ corresponds to the value of one of the quasi-degenerate masses. The effective Majorana mass parameter, $m_{\beta \beta}$, is a combination of the three neutrino masses, oscillation parameters, and Majorana phases. In the standard case of three Majorana neutrinos, this sets an upper bound on the minimal neutrino mass.}. Most relevant for cosmology are the neutrino oscillation experiments, which are sensitive to the mass squared differences. Global fits to the experimental data indicate that $\Delta m_{21}^2\equiv m_2^2-m_1^2\sim\SI{7.5e-5}{\eV^2}$ and $\rvert\Delta m^2_{31}\rvert\!\equiv\,\rvert m_3^2 - m_1^2\rvert\sim\SI{2.5e-3}{\eV^2}$ \citep{esteban20,nufit24,capozzi21,desalas21,pdg22}. These set a lower limit on the sum of the three neutrino masses, depending on the sign of $\Delta m^2_{31}$:
\begin{align*}
    \textstyle\sum m_\nu\!=m_1 + \sqrt{m_1^2 + \Delta m^2_{21}} + \sqrt{m_1^2+\Delta m^2_{31}} > \SI{0.059}{\eV},
\end{align*}

\noindent
if $m_1<m_2<m_3$ (the normal ordering) or
\begin{align*}
    \textstyle\sum m_\nu\!=m_3 + \sqrt{m_3^2-\Delta m^2_{31}} + \sqrt{m_3^2-\Delta m^2_{31}+\Delta m^2_{21}} > \SI{0.10}{\eV},
\end{align*}

\noindent
if $m_3<m_1<m_2$ (the inverted ordering).

Recent cosmological constraints are approaching or breaching these lower limits, when assuming a standard $\Lambda$CDM model. The strongest bounds generally result from the combination of \emph{Planck} measurements of the cosmic microwave background (CMB) \cite{planck18} and different probes of the large-scale structure \cite{palanque20,divalentino21,brieden22}. With the latest baryon acoustic oscillations (BAO) and full-shape measurements by DESI \citep{desi24a,desi24b,desi24c,desi_kp7b}, the limit has tightened to $\sum m_\nu<\SI{0.071}{\eV}$ (95\%) \citep{desi24b}. Incorporating also other background measurements, a bound of $\sum m_\nu<\SI{0.043}{\eV}$ (95\%) was obtained \citep{wang24}, in significant tension with the lower bounds for both the normal and inverted orderings.

The tension with neutrino oscillations motivates the search for alternative cosmological models. Alternative models of dark energy are of particular interest, due to the degeneracy between the sum of neutrino masses, $\sum m_\nu$, and the dark energy equation of state, $w$ \cite{hannestad05,lorenz17,choudhury18,vagnozzi18,upadhye19,liu20,choudhury20,sharma22,yadav24}. Recently, DESI reported hints of an evolving dark energy equation of state of $2.5-3.9\sigma$ \cite{desi24b,lodha24,calderon24}. In the $w_0w_a$CDM model, the equation of state is a function of the expansion factor, $w(a) = w_0 + w_a(1-a)$. This simple empirical model is flexible enough to match the observational predictions for DESI of physically motivated dark energy models \cite{chevallier01,linder03,deputter08}. For instance, the mirage class of dark energy models, in which the equation of state crosses $w\approx-1$ around $z\approx0.4$ and therefore resembles that of a cosmological constant, and which provide an improved fit to the DESI data compared to $\Lambda$CDM, can be described by $w_a=-3.66(1+w_0)$ \cite{linder07,lodha24}. When adopting the general $w_0w_a$CDM model, the DESI constraint on the neutrino masses is relaxed to $\sum m_\nu<\SI{0.195}{\eV}$ (95\%), which is consistent with neutrino oscillation data.

The possibility of systematic errors must also be considered. One concern is the presence of an oscillatory residual in the best-fitting $\Lambda$CDM temperature power spectrum from \emph{Planck} \cite{planck20}. Such a feature could be explained by additional smoothing of the acoustic peaks, which is a characteristic signature of gravitational lensing. The anomaly can be quantified by scaling the gravitational lensing potential by a factor $A_\text{lens}$, defined such that $A_\text{lens}=1$ in the absence of systematics or new physics \cite{calabrese08,renzi18,mokeddem23}. Various analyses have found a preference for $A_\text{lens}>1$ \cite{planck18,divalentino20,divalentino20b,ballardini22,mokeddem23}, albeit with reduced significance since the latest data release of \emph{Planck} (PR4) \cite{efstathiou19,tristram24}. Since massive neutrinos suppress the growth of cosmic structure and thereby the strength of lensing, a preference for $A_\text{lens}>1$ implies tight upper limits on the neutrino mass \cite{mccarthy18,choudhury20,divalentino20,elbers20,allali24,craig24}, and could even be interpreted as $\sum m_\nu < 0$ \cite{elbers20,allali24,craig24}.

Nearly all cosmological analyses to date have imposed the physical constraint, $\sum m_\nu\geq 0$, and found that the marginal posterior distribution, $P(\sum m_\nu)$, peaks at $\sum m_\nu=0$. As a consequence, imposing more restrictive constraints from neutrino oscillations leads to outcomes that are dominated by those constraints. Indeed, imposing the prior, $\sum m_\nu>\SI{0.059}{\eV}$, significantly degrades the DESI constraint to $\sum m_\nu<\SI{0.113}{\eV}$ (95\%) \citep{desi24b}. This dependence on \emph{a priori} assumptions also calls into question the $\sum m_\nu\geq 0$ prior. Only by relaxing this constraint, can we assess whether cosmological data are compatible with physical neutrino masses and reveal the dependence of the central value on the data and choice of model. This is necessary to determine whether changes in the upper bound are due to increased precision or shifts in the posterior, which are otherwise easily confused.

Previously, using frequentist methods, \cite{planck14} extrapolated a parabolic fit to the profile likelihood curve and estimated a minimum at $\sum m_\nu=-0.05\pm0.15\,\si{\eV}$ (68\%) from \emph{Planck} and BAO data. From the Bayesian point of view, \cite{alam21} extrapolated the marginal posterior distribution to negative values by fitting a Gaussian distribution to $P(\sum m_\nu)$ and obtained $\sum m_\nu=-0.026\pm0.074\,\si{\eV}$ (68\%) from \emph{Planck} and SDSS BAO. More recently, \cite{craig24} extended the analysis to negative neutrino masses by expressing the effect of neutrinos on the CMB in terms of $A_\text{lens}(\sum m_\nu)$, finding $\sum m_\nu=-0.16\pm0.09\,\si{\eV}$ (68\%) from \emph{Planck}, ACT, and DESI. However, these approaches cannot fully characterize the effects of $\sum m_\nu < 0$, such as the impact on the expansion history, or capture parameter correlations independently of $A_\text{lens}$. In this paper, we introduce a model that extends consistently the domain to negative masses for all observables and examine the preference for $\sum m_\nu <0$.

\section{Negative neutrino masses}

Formally, the Friedmann equations that govern the expansion of space depend only on the neutrino masses squared, $m_i^2$, via expressions like
\begin{align}
   \Omega_{\nu}(a) = \sum_{i=1}^{N_\nu} \frac{8 G T_\nu^4}{3\pi H_0^2}\int_0^\infty\frac{x^2\sqrt{x^2+a^2m_i^2/T_\nu^2}}{1+e^x}\mathrm{d}x, \label{eq:Omega_nu}
\end{align}

\noindent
where $N_\nu$ is the number of neutrino species, $T_\nu$ the present-day neutrino temperature, and $H_0$ the Hubble constant. Only at late times, when the neutrinos become non-relativistic and the $a^2m_i^2/T_\nu^2$ term dominates, does this expression reduce to the well-known approximation
\begin{align}
   \Omega_\nu \approx \frac{\textstyle\sum m_\nu}{93.14h^2}. \label{eq:Onu_approximation}
\end{align}
The effect of Eq.~\eqref{eq:Omega_nu} is to produce a greater radiation density while neutrinos are relativistic and a greater matter density once neutrinos become non-relativistic. Without attributing the effect to neutrinos, a phenomenological term of the form, $\Omega_{\nu,\text{eff}}(a) = \kappa \Omega_\nu(a)$, would behave in the opposite way for $\kappa < 0$, reducing the radiation density at early times and the matter density at late times. To make contact with terrestrial constraints, we could reinterpret such a term as an \emph{effective neutrino mass} parameter, $\sum m_{\nu,\text{eff}}\equiv 93.14h^2\, \Omega_{\nu,\text{eff}}$ \footnote{In theoretical particle physics, fermion masses may carry a complex phase and be negative, depending on their CP properties. However, we reiterate that it is the energy (depending on the absolute value of the mass) that enters the Friedmann equations. Hence, negative neutrino masses in cosmology more accurately correspond to negative energies. We use the term effective to emphasize that negative values are unphysical and rather signal the presence of systematic errors, a non-standard cosmological evolution or new neutrino properties.}.

\begin{figure}
    \includegraphics{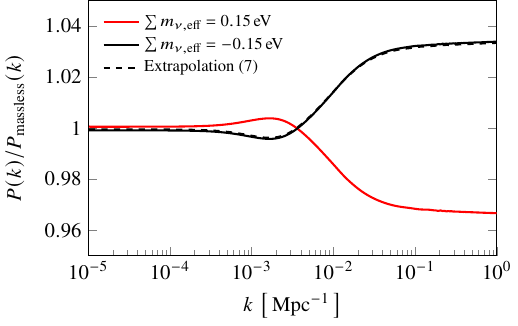}\vspace{-1em}
    \caption{The effects of positive and negative effective neutrino masses on the dark matter power spectrum at $z=0$. We show both the exact linear theory calculation for $\SI{0.15}{\eV}$ (solid red) and $\SI{-0.15}{\eV}$ (solid black), as well as the linear extrapolation \eqref{eq:nu_eff_def} in dashed black, which agrees with the exact calculation to within $0.1\%$.}
    \label{fig:pk_validation}
\end{figure}

We can extend this behaviour beyond the Friedmann equations to all orders in cosmological perturbation theory by consistently replacing the neutrino energy, $\epsilon$, with an effective neutrino energy,
\begin{align}
    \epsilon_\text{eff}=\text{sgn}(m_\nu)\sqrt{x^2 + a^2m_\nu^2/T_\nu^2},
\end{align}

\noindent
where $\text{sgn}$ is the sign function. For instance, the familiar first-order equations for the multipole moments, $\Psi_\ell$, of the neutrino distribution function in conformal Newtonian gauge \cite{ma95} become
\begin{align}
    \dot{\Psi}_0 &= -\frac{qk}{\epsilon_\text{eff}}\Psi_1 - \dot{\phi}\frac{\mathrm{d}\ln \bar{f}}{\mathrm{d}\ln q}, \label{eq:b0}\\
    \dot{\Psi}_1 &= \frac{qk}{3\epsilon_\text{eff}}(\Psi_0 - 2\Psi_2) - \frac{\epsilon_\text{eff}k}{3q}\psi\frac{\mathrm{d}\ln \bar{f}}{\mathrm{d}\ln q}, \label{eq:b1}\\
    \dot{\Psi}_\ell &= \frac{qk}{(2\ell+1)\epsilon_\text{eff}}\left[\ell\Psi_{\ell-1} - (\ell+1)\Psi_{\ell+1}\right], \;\;\; (\ell\geq2), \label{eq:b2}
\end{align}

\noindent
where dots denote conformal time derivatives, $\phi$ and $\psi$ are the metric perturbations, $k$ is the wavenumber, $q$ the neutrino momentum, and $\bar{f}$ the unperturbed distribution function.

The following operational definition offers a simpler calculation. We restrict attention to the case where all masses are either positive or negative. For any cosmological observable $X$, such as the CMB temperature power spectrum, $X=C^\text{TT}_{\ell}$, and a set of fixed parameters, $\bm{\theta}=\{\omega_\text{b},\omega_\text{c},\theta_{\text{s}},\tau,A_s,n_s\}$, we define the prediction for the effective mass, $\sum m_{\nu,\text{eff}}$, as
\begin{align}
    \!\!X_{\bm{\theta}}^{\sum\! m_{\nu,\text{eff}}} \!\equiv\! X_{\bm{\theta}}^{\sum\! m_{\nu}=0} \!+ \text{sgn}(\textstyle\sum m_{\nu,\text{eff}}) \left[X_{\bm{\theta}}^{\rvert\sum\! m_{\nu,\text{eff}}\rvert} \!-\! X_{\bm{\theta}}^{\sum\! m_{\nu}=0} \right], \label{eq:nu_eff_def}
\end{align}

\noindent
where $X_{\bm{\theta}}^{\rvert\sum\! m_{\nu,\text{eff}}\rvert}$ is the prediction for positive neutrino masses $\rvert m_i\rvert$. We implemented both the exact approach and the extrapolation \eqref{eq:nu_eff_def} in the \texttt{CLASS} \citep{lesgourgues11,lesgourgues11b} and \texttt{cobaya} \citep{torrado21} codes. The two methods agree to excellent precision, as shown in Fig.~\ref{fig:pk_validation} for the effect on the dark matter power spectrum. The results presented in this paper were obtained using \eqref{eq:nu_eff_def}, but we verified that this does not affect the conclusions.

A key advantage of our approach is that one recovers exactly the physical neutrino model for $\sum m_{\nu,\text{eff}} = \sum m_\nu \geq 0$. Moreover, we no longer need to assume a Gaussian functional form for the marginalized posterior distribution, nor make any assumptions about parameter correlations. We apply our model to the latest BAO measurements from DESI \cite{desi24a,desi24c} and CMB temperature and polarization measurements from \emph{Planck}, using the low-$\ell$ \texttt{Commander} and \texttt{SimAll} likelihoods \cite{planck20} and the high-$\ell$ \texttt{CamSpec} likelihood \citep{efstathiou19,rosenberg22} based on the PR4 release. In some cases, we also include CMB lensing measurements based on ACT DR6 \cite{madhavacheril24,qu24,maccrann24} and \emph{Planck} PR4 \cite{carron22}. Our primary analysis is explicitly blind to the constraints from neutrino oscillations. In this case, we assume a degenerate mass spectrum with $N_\nu=3$ species and $m_{\nu,\text{eff}}\equiv m_1=m_2=m_3$, with a uniform prior, $m_{\nu,\text{eff}}\in[-1.5,1.5]\,\si{\eV}$. When we perform a combined analysis of cosmological data and laboratory constraints, we adopt Gaussian likelihoods on $\Delta m^2_{21}$ and $\rvert\Delta m^2_{31}\rvert$, based on global fits to the experimental data \cite{pdg22}, and fix $m_1,m_2,m_3$ in terms of $\Delta m^2_{21}, \rvert\Delta m^2_{31}\rvert,\beta,$ and $m_\text{lightest}$, where $\beta$ is a binary variable for the mass ordering and $m_\text{lightest}$ the lightest neutrino mass \cite{loureiro19}. We then adopt a uniform prior on $m_\text{lightest}\in[0,0.5]\,\si{\eV}$.

\begin{figure}
    \includegraphics{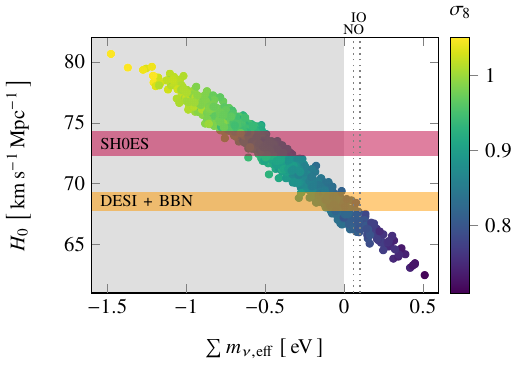}\vspace{-0.5em}
    \caption{Constraints on the effective neutrino mass, $\sum m_{\nu,\text{eff}}$, the Hubble constant, $H_0$, and the amplitude of matter fluctuations, $\sigma_8$, from \emph{Planck} temperature and polarization data \cite{planck20,efstathiou19,rosenberg22}, assuming $\Lambda$CDM. The degeneracy between these parameters can be broken with measurements of the expansion history. Shown are the $\pm1\sigma$ bounds from DESI BAO, combined with a Big Bang nucleosynthesis prior on $\Omega_\text{b}h^2$ \cite{desi24a,desi24c}, and the SH0ES measurement of $H_0$ from the local distance ladder \cite{ries22}. The unphysical regime, $\sum m_{\nu,\text{eff}}<0$, is grey and the dotted lines indicate the lower bounds from neutrino oscillations for the normal ordering (NO) and inverted ordering (IO).}
    \label{fig:planck_posterior}
\end{figure}

\begin{figure*}
    \subfloat{
        \includegraphics{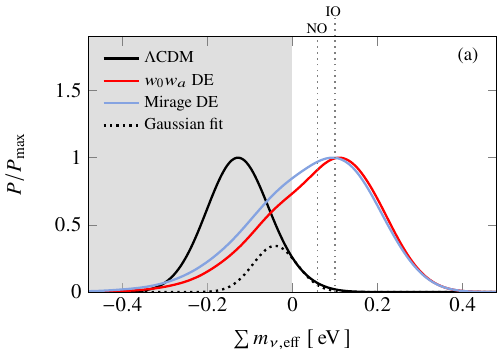}
        \label{fig:mnu_posterior}
    }
    \subfloat{
        \includegraphics{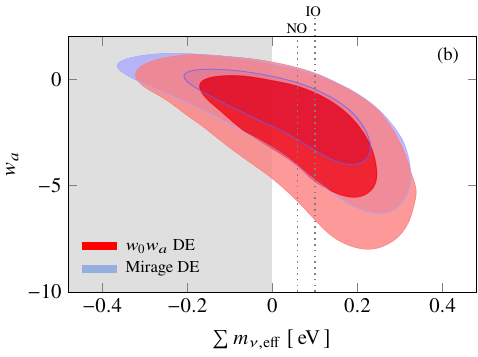}
        \label{fig:contours}
    }\vspace{-0.5em}
    \caption{(a) Posterior distribution of $\sum m_{\nu,\text{eff}}$ from \emph{Planck}, ACT, and DESI data, for $\Lambda$CDM, $w_0w_a$CDM, and mirage dark energy. The black dotted line is a Gaussian fit to the $\Lambda$CDM posterior restricted to $\sum m_{\nu,\text{eff}}\geq0$, normalized by the same $P_\text{max}$. (b) The 68\% and 95\% constraints on $\sum m_{\nu,\text{eff}}$ and $w_a$ for the same models and data.}
    \label{fig:main_constraints}
\end{figure*}

\section{Results}

In the first instance, we assume $\Lambda$CDM and only use \emph{Planck} CMB temperature and polarization data. The resulting constraints on $\sum m_{\nu,\text{eff}}$, and the Hubble constant, $H_0$, are shown in Fig.~\ref{fig:planck_posterior}, where the colours indicate the amplitude of matter fluctuations, $\sigma_8$. The results are consistent with the lower bounds from neutrino oscillations, which are shown as vertical dotted lines, at the $1.9\sigma$ level for the normal ordering and at $2.1\sigma$ for the inverted ordering \footnote{We verified that the level of tension is robust against projection effects \cite{noriega24} by also computing the $\Delta\chi^2$ between the best-fitting models with and without laboratory constraints.}. The figure clearly demonstrates the geometric degeneracy between $\sum m_{\nu,\text{eff}}$ and $H_0$. This degeneracy can be broken with measurements of the expansion history. We show two such measurements. In both cases, the preference for $\sum m_{\nu,\text{eff}}<0$ increases compared to the CMB-only case. The first is the SH0ES measurement of $H_0=73.04\pm1.04\,\si{\km\per\s\per\Mpc}$ from Cepheid variable stars and type 1a supernovae \cite{ries22}. Reconciling the SH0ES measurement with \emph{Planck} requires a negative effective neutrino mass of $\sum m_{\nu,\text{eff}}=-0.5\pm0.1\,\si{\eV}$ and values of $\sigma_8=0.92\pm0.02$ that are large compared to clustering measurements \cite{hamana20,heymans21,abbott22,brieden22,kramstra24}.

For illustrative purposes, we also show the determination of $H_0=68.53\pm0.80\,\si{\km\per\s\per\Mpc}$ from DESI BAO combined with a Big Bang nucleosynthesis prior on $\Omega_\text{b}h^2$ \cite{desi24b}. When performing a formal analysis of \emph{Planck} CMB temperature and polarization and DESI BAO data, we find that the tension with the lower bound from neutrino oscillations increases from $1.9\sigma$ to $2.7\sigma$ for the normal ordering and from $2.1\sigma$ to $3.2\sigma$ for the inverted ordering. This trend continues with the addition of CMB lensing measurements from \emph{Planck} and ACT DR6 \cite{carron22,madhavacheril24,qu24,maccrann24}, increasing the tension to $2.8\sigma$ and $3.3\sigma$. The resulting marginalized posterior obtained from \emph{Planck} + ACT + DESI is shown as a black line in Fig.~\ref{fig:mnu_posterior}. As a point of reference, we compare our results with those obtained for a Gaussian extrapolation of $P(\sum m_\nu)$ to $\sum m_\nu<0$, as used previously in \cite{alam21,allali24}. We find that the Gaussian procedure significantly underestimates the preference for negative effective neutrino masses, as can be seen from the dotted line in Fig.~\ref{fig:mnu_posterior}. The Gaussian fit yields a central value of $\sum m_{\nu,\text{eff}}=-0.041\pm0.052\,\si{\eV}$ (68\%), compared to the full posterior mean of $\sum m_{\nu,\text{eff}}=-0.125_{-0.070}^{+0.058}\,\si{\eV}$ (68\%).


\begin{figure*}
    \subfloat{
        \includegraphics{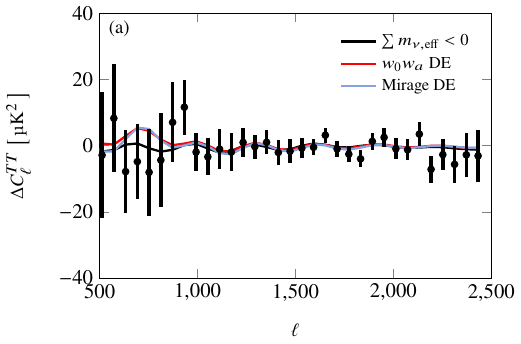}
        \label{fig:cmb_bao_plot_a}
    }
    \subfloat{
        \includegraphics{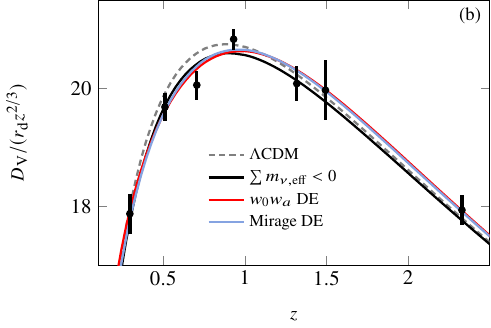}
        \label{fig:cmb_bao_plot_b}
    }\vspace{-0.5em}
    \caption{(a) Residuals of the best-fitting $\Lambda$CDM temperature power spectrum with physical neutrino masses, together with the best-fitting $\Lambda$CDM model with $\sum m_{\nu,\text{eff}}<0$, and the best-fitting $w_0w_a$CDM and mirage models with physical neutrino masses. (b) DESI measurements of the angle-averaged distance ratio, $D_\text{V}/r_\text{d}$, arbitrarily rescaled by a factor of $z^{2/3}$ for clarity, together with the predictions for the best-fitting $\Lambda$CDM, $w_0w_a$CDM, and mirage models with physical neutrino masses and the best-fitting $\Lambda$CDM model with $\sum m_{\nu,\text{eff}}<0$.}
    \label{fig:cmb_bao_plot}
\end{figure*}

Motivated by the hint of evolving dark energy reported by DESI \cite{desi24b}, we repeat the analysis for the $w_0w_a$CDM model for \emph{Planck}, ACT, and DESI data. We obtain a posterior mean of $\sum m_{\nu,\text{eff}}=0.06_{-0.10}^{+0.15}\,\si{\eV}$ (68\%) and an upper bound of $\sum m_{\nu,\text{eff}}<\SI{0.24}{\eV}$ (95\%). Hence, the data are now fully consistent with the lower bounds from neutrino oscillations. The full posterior is shown as a red line in Fig.~\ref{fig:mnu_posterior}. Interestingly, the reduction in tension is driven not just by an increase in uncertainty, from $\sigma(\sum m_{\nu,\text{eff}})=0.07\,\si{\eV}$ for $\Lambda$CDM to $\sigma(\sum m_{\nu,\text{eff}})=0.13\,\si{\eV}$, but primarily by a large shift in the central value of $\Delta(\sum m_{\nu,\text{eff}})=\SI{0.24}{\eV}$. This shift can only be quantified by allowing negative effective neutrino masses. A similar shift is observed in the CMB-only case, in the absence of DESI BAO data. We also consider the mirage class of dark energy models and show the resulting posterior as a blue line. We obtain $\sum m_{\nu,\text{eff}}=0.04_{-0.11}^{+0.15}\,\si{\eV}$ (68\%) and an upper bound of $\sum m_{\nu,\text{eff}}<\SI{0.24}{\eV}$ (95\%), similar to the $w_0w_a$CDM case. In Fig.~\ref{fig:contours}, we show the 68\% and 95\% constraints in the plane of $\sum m_{\nu,\text{eff}}$ and $w_a$. The dark energy equation of state parameters are degenerate with the sum of neutrino masses, with larger neutrino masses requiring $w_a<0$. We note that an evolving dark energy equation of state appears to be necessary. Within the $w$CDM model, with a constant equation of state, $w$, we obtain $w=-1.01\pm0.08$, consistent with $\Lambda$CDM, and $\sum m_{\nu,\text{eff}}=-0.12\pm0.10\,\si{\eV}$ (68\%). 

To understand these results, we show the residuals of the CMB temperature power spectrum, relative to the best-fitting $\Lambda$CDM model with physical neutrino masses, obtained from a combined analysis of cosmological data (\emph{Planck} + ACT + DESI) and laboratory constraints, in Fig.~\ref{fig:cmb_bao_plot_a}. There is a clear oscillatory feature that gives rise to a preference for $A_\text{lens}>1$. The same oscillations can also be described by abandoning the laboratory constraints and allowing $\sum m_{\nu,\text{eff}}<0$, since negative effective neutrino masses enhance the growth of density perturbations and boost the lensing potential. Gravitational lensing is similarly enhanced for the best-fitting dark energy models with physical neutrino masses (both for $w_0w_a$ and mirage dark energy), obtained from a combined analysis of cosmological data and laboratory constraints. The preference for these models is not just driven by the CMB, but also by the expansion history. Fig.~\ref{fig:cmb_bao_plot_b} shows the isotropic BAO distance measurements from DESI, along with the best-fitting $\Lambda$CDM model with negative neutrino masses (black) and the best-fitting models with physical neutrino masses for $\Lambda$CDM (grey dashed), $w_0w_a$ (red) and mirage dark energy (blue). Within $\Lambda$CDM, adopting $\sum m_{\nu,\text{eff}}<0$ provides a better fit to the BAO data, especially at $z<1$. The evolving dark energy models fit the data even better, but with one or two additional parameters. The $\chi^2$ for these four models are given in Table~\ref{tab:chi_squared}, confirming that negative values of effective neutrino masses and evolving dark energy both improve the fit to the BAO data, the high-$\ell$ CMB temperature data, and the low-$\ell$ CMB polarization data.

\begin{figure}
    \includegraphics{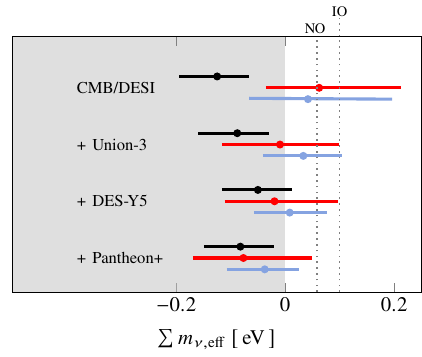}\vspace{-0.5em}
    \caption{The 68\% constraints on the effective neutrino mass, $\sum m_{\nu,\text{eff}}$, from CMB/DESI (\emph{Planck} + ACT + DESI) alone and combined with supernovae from Union-3, DES-Y5, and Pantheon+.}
    \label{fig:whiskers}
\end{figure}

Finally, we consider the impact of constraints on the expansion history from type 1a supernovae. The inclusion of supernovae helps to constrain the dark energy parameters by breaking the degeneracy between $w_0$ and $w_a$ \cite{desi24b}, but also leads to weaker constraints on $\sum m_\nu$ \cite{allali24}. We consider three different data sets: Union-3 \cite{rubin23}, Dark Energy Survey (DES) Year 5 \cite{des24}, and Pantheon+ \cite{scolnic22,brout22}. These data sets have many supernovae in common but differ in their treatment of systematic errors. We show the 68\% constraints on $\sum m_{\nu,\text{eff}}$ from \emph{Planck} + ACT + DESI alone and combined with each of the three supernova data sets in Fig.~\ref{fig:whiskers}. The inclusion of supernovae helps to constrain $H_0$, which for $\Lambda$CDM shifts $\sum m_{\nu,\text{eff}}$ to slightly larger values along the geometric degeneracy. Nevertheless, the tension with the lower bound for the normal ordering remains at $2.4\sigma$ (\mbox{Union-3}), $2.0\sigma$ (DES-Y5), and $2.3\sigma$ (\mbox{Pantheon+}). For $w_0w_a$CDM, the values of $\sum m_{\nu,\text{eff}}$ are reduced, but remain compatible with the lower bound for the normal ordering. In general, the greater the pull away from $\Lambda$CDM (with $w_0=-1$ and $w_a=0$), the greater the shift in $\sum m_{\nu,\text{eff}}$ between $\Lambda$CDM and $w_0w_a$CDM. Hence, the shift is most pronounced for \mbox{Union-3} ($w_0=-0.69_{-0.12}^{+0.10}$, $w_a=-1.04_{-0.44}^{+0.54}$) and least significant for Pantheon+ ($w_0=-0.866_{-0.069}^{+0.062}$, $w_a=-0.43_{-0.30}^{+0.38}$). This confirms that a large shift away from $\Lambda$CDM is needed to reconcile cosmology with positive neutrino masses satisfying the constraints from neutrino oscillations. For the mirage dark energy model, the results with supernovae are remarkably consistent with \emph{Planck} + ACT + DESI alone, but with greatly reduced uncertainty. Moreover, once combined with supernovae, mirage dark energy favours the largest neutrino masses, which are always compatible with the lower bound for the normal ordering \footnote{As we prepared this manuscript, Ref.~\cite{green24} presented an analysis with negative neutrino masses, reaching different conclusions about dark energy and the expansion history. These differences warrant further investigation.}.

\begin{table}
\caption{\label{tab:chi_squared}
Values of $\chi^2$ for the best-fitting $\Lambda$CDM model with physical neutrino masses, together with the $\Delta\chi^2$ for the best-fitting $\Lambda$CDM model with negative effective neutrino masses and the best-fitting $w_0w_a$ and mirage DE models with physical neutrino masses.}
\begin{ruledtabular}
\begin{tabular}{lcccccc}
 & & \multicolumn{2}{c}{high-$\ell$ CMB} & \multicolumn{2}{c}{low-$\ell$ CMB} & CMB\\
 & BAO & TT & TEEE & TT & EE & Lensing\\
\hline
$\Lambda$CDM & $14.93$ & $6419.56$ & $4121.90$ & $22.73$ & $398.27$ & $19.88$\\
\hline
$\sum m_{\nu,\text{eff}}<0$ & $-2.13$ & $-2.09$ & $-0.25$ & $+0.55$ & $-2.55$ & $+0.17$\\
$w_0w_a$ DE & $-4.13$ & $-3.13$ & $+0.47$ & $+1.56$ & $-2.57$ & $-0.15$\\
Mirage DE & $-3.28$ & $-2.71$ & $+0.65$ & $+1.23$ & $-2.57$ & $-0.19$\\
\end{tabular}
\end{ruledtabular}
\end{table}

\section{Conclusion}

Cosmological constraints on the neutrino mass sum, $\sum m_\nu$, are becoming increasingly sensitive to prior assumptions about the mass spectrum. In this paper, we showed how the domain of an effective neutrino mass parameter, $\sum m_{\nu,\text{eff}}$, can be consistently extended to negative values. By abandoning the physical constraint, $\sum m_\nu\geq0$, and adopting an effective mass parameter with a broad prior, we can assess whether cosmological data are compatible with laboratory constraints, determine the sensitivity of the data independently of the prior, and reveal how the central value depends on the data and choice of model. Analyzing cosmological data from \emph{Planck}, ACT, and DESI in the context of the $\Lambda$CDM model, we found a preference for negative masses and a tension of $2.8-3.3\sigma$ with the lower bounds on $\sum m_\nu$ that apply in the case of positive neutrino masses satisfying the constraints from neutrino oscillations.

We showed that adopting an evolving dark energy equation of state, combined with physical neutrino masses, leads to similar predictions as the $\Lambda$CDM model with $\sum m_{\nu,\text{eff}}<0$. In particular, both models predict additional gravitational lensing of the CMB. As a result, evolving dark energy models can address both the preference for additional CMB lensing and the tension with neutrino oscillations. The mirage class of dark energy models appears promising \cite{linder07,lodha24}, favouring larger neutrino masses compatible with laboratory data, with similar uncertainty as $\Lambda$CDM once combined with supernova data. The lower bounds from neutrino oscillations are saturated in the case where the lightest neutrino is massless. Measurement of a non-zero lightest neutrino mass by KATRIN \cite{katrin24}, Project 8 \cite{esfahani17}, or from neutrinoless double-$\beta$ decay \cite{nemo17,adams21,abe24} would further challenge the assumptions of a standard cosmological evolution and cosmic neutrino background.

\begin{acknowledgments}
WE acknowledges useful discussions with the DESI CPE working group, thanking in particular Eva-Maria Mueller for suggesting the Gaussian comparison. WE, CSF, AJ, and BL acknowledge STFC Consolidated Grant ST/X001075/1 and support from the European Research Council through ERC Advanced Investigator grant, DMIDAS [GA 786910] to CSF. This work used the DiRAC@Durham facility managed by the Institute for Computational Cosmology on behalf of the STFC DiRAC HPC Facility (www.dirac.ac.uk). The equipment was funded by BEIS capital funding via STFC capital grants ST/K00042X/1, ST/P002293/1 and ST/R002371/1, Durham University and STFC operations grant ST/R000832/1. DiRAC is part of the National e-Infrastructure.
\end{acknowledgments}

\appendix

\bibliography{main}

\end{document}